\documentclass{article}
\usepackage{icmc2025_paper_template}
\usepackage{times}
\usepackage{ifpdf}
\usepackage{soul}
\usepackage[english]{babel}

\usepackage{booktabs}
\usepackage{amsmath,amssymb}
\usepackage{amsfonts}       

\def\bm{\boldsymbol}
\def\Z{\mathbb Z}
\def\R{\mathbb R}
\def\Q{\mathbb Q}
\def\ord{\mathop{\mathrm{ord}}}
\def\bvec{\mathop{\mathrm{bvec}}}
\def\vvec{\mathop{\mathrm{vec}}}
\def\Id{\mathop{\mathrm{Id}}}

\def\papertitle{Music102: An $D_{12}$-equivariant transformer for chord progression accompaniment}
\def\firstauthor{Weiliang Luo}
\def\secondauthor{Second Author}
\def\thirdauthor{Third Author}

\newif\ifpdf
\ifx\pdfoutput\relax
\else
   \ifcase\pdfoutput
      \pdffalse
   \else
      \pdftrue
  \fi
\fi

\ifpdf 
  \usepackage[pdftex,
    pdftitle={\papertitle},
    pdfauthor={\firstauthor, \secondauthor, \thirdauthor},
    bookmarksnumbered, 
    pdfstartview=XYZ 
   ]{hyperref}

  \usepackage[pdftex]{graphicx}
  \graphicspath{{./figures/}}
  \DeclareGraphicsExtensions{.pdf,.jpeg,.png}

  \usepackage[figure,table]{hypcap}

\else 
  \usepackage[dvips,
    bookmarksnumbered, 
    pdfstartview=XYZ 
  ]{hyperref}  

  \usepackage[dvips]{epsfig,graphicx}
  \graphicspath{{./figures/}}
  \DeclareGraphicsExtensions{.eps}

  \usepackage[figure,table]{hypcap}
\fi

\hypersetup{
    colorlinks,%
    citecolor=black,%
    filecolor=black,%
    linkcolor=black,%
    urlcolor=black
}

\title{\papertitle}

\oneauthor
  {\firstauthor} {Massachusetts Institute of Technology\\ %
    {\tt \href{anonymous@myorg.org}{luowl7@mit.edu}}}

\begin{document}
\capstartfalse
\maketitle
\capstarttrue
\begin{abstract}
We present Music102, an advanced model aimed at enhancing chord progression accompaniment through a $D_{12}$-equivariant transformer. Inspired by group theory and symbolic music structures, Music102 leverages musical symmetry--such as transposition and reflection operations--integrating these properties into the transformer architecture. By encoding prior music knowledge, the model maintains equivariance across both melody and chord sequences. The POP909 dataset was employed to train and evaluate Music102, revealing significant improvements over the non-equivariant Music101 prototype Music101 in both weighted loss and exact accuracy metrics, despite using fewer parameters. This work showcases the adaptability of self-attention mechanisms and layer normalization to the discrete musical domain, addressing challenges in computational music analysis. With its stable and flexible neural framework, Music102 sets the stage for further exploration in equivariant music generation and computational composition tools, bridging mathematical theory with practical music performance.
\end{abstract}

\section{Introduction}\label{sec:introduction}
In the burgeoning age of AI arts, generative AI, represented by the diffusion model, has been profoundly influencing the concept of digital paintings, while music production remains a frontier for machine intelligence to explore. For an AI composer, there's a long way to achieve the holy grail of creating a complete classical symphony; however, simpler tasks such as fragment generation and accompaniment are within our reach. Inspired by the wide demand for music education and daily music practice, we designed a prototypical model, Music101, which succeeded in predicting a reasonable chord progression given the single-track melody of pop music. Although the model was observed to have captured some essential music patterns, its performance was still far from a real application. The quantity and the quality of the available dataset are quite limited, so any complication of the architecture didn't benefit the previous experiments. 

Nevertheless, there are hidden inductive biases in this task to be leveraged. Between music and mathematics is a natural connection, especially in the language of symbolic music. With equal temperament, transposition, and reflection symmetry emerge from the pitch classes on the keyboard. The conscious application of group theory on music dates back to Schoenberg's atonal music, and the relevant group actions have long been in the toolbox of classical composers, even earlier than Bach. In this work, we proposed Music102, which encodes prior knowledge of music by leveraging this symmetry in the music representation. 

\section{Related work}\label{sec:related}
The notion of the symmetry of pitch classes is fundamental in music theory \cite{mazzola2012topos}, where the group theory exerts its power as in spatial objects \cite{papadopoulos2014mathematics}. As an essential prior knowledge of the music structure, computational music studies have embraced it in various tasks, such as transposition-invariant music metric \cite{hadjeres2017deep}, transposition-equivariant pitch estimation \cite{riou2023pesto,cwitkowitz2024toward}. Some work tried to extract this structure from music pieces in priori as well \cite{lostanlen2020learning}.

Between notes and words, there is an analogy between natural languages and music. Therefore, popular frameworks for natural language tasks, especially the transformer, are proven indispensable in symbolic music understanding and generation with the awareness of inherent characteristics of music, like the long-term dependence and timescale separation explored in Music transformer \cite{huang2018music} and Museformer \cite{yu2022museformer}. The triumphs in the time domain encourage us to focus on the rich structures in the frequency domain, that is, the pitch classes in symbolic music.

The introduction of equivariance into attentive layers has been given much attention in building expressive and flexible models for grid, sequence, and graph targets. SE(3)-transformer \cite{fuchs2020se} provides a good example of how the attention mechanism works naturally with invariant attention scores and equivariant value tensors. The non-linearity and layer normalization layer in this work is inspired by Equiformer \cite{liao2022equiformer,liao2023equiformerv2} and the implementation of $S^2$ activation in E3NN \cite{geiger2022e3nn}.

\section{Background}
\subsection{Equal temperament}
The notion of a sound's \textbf{pitch} is physically instantiated by the frequency of its vibration.  Due to the physiological features of human ears and/or centuries of cultural construction, sounds of frequencies with a simple integer ratio make people feel harmonious. The relation between two pitches with the simplest non-trivial ratio, $1:2$, is called an \textbf{octave}, the best harmony we can create. Based on the octave, pitches with a frequency ratio of $1:2^n\ (n\in\Z)$ build an equivalence relation, which partitions the set of pitches into \textbf{pitch classes}.

However, all frequencies are neither implemented in music instruments nor allowed in most music compositions. A \textbf{temperament} is a criterion for how people define \textit{the relationship between} legal pitches among available frequencies. One of the most essential topics of the temperament is how to make it finer-grained to accommodate useful pitches. The \textbf{equal temperament} is the most popular scheme in modern music. It subdivides an octave into 12 minimal intervals, called \textbf{semitones} or \textbf{keys}, each of which is a multiple of $2^{1/12}$ on the frequency. Their equality is manifested on the logarithm scale of the frequency. Starting from a pitch class $\mathrm C$, a set of 12 pitch classes $\mathcal P$ spanning the octave is iteratively defined by the semitone. Each of their \textbf{pitch class name} as a letter $\mathrm{P}$ and its ordinal number $\ord(\mathrm{P})$ in the frequency ascending order are in~\autoref{tab:octave}. Some pitch classes have two equivalent names in equal temperament, with preference only when discussing the interaction between pitch classes.

\begin{table}[!htbp]
    \centering
    \footnotesize
    \begin{tabular}{c|cccccc}
    \toprule
    $\mathrm{P}$ & $\mathrm{C}$ & $\mathrm{C}^\sharp/\mathrm{D}^\flat$ & $\mathrm{D}$ & $\mathrm{D}^\sharp/\mathrm{E}^\flat$ & $\mathrm{E}$ & $\mathrm{F}$ \\
    \midrule
    $\ord(\mathrm{P})$ & 0 & 1 & 2 & 3 & 4 & 5\\
    \toprule
    $\mathrm{P}$ & $\mathrm{F}^\sharp/\mathrm{G}^\flat$ & $\mathrm{G}$ & $\mathrm{G}^\sharp/\mathrm{A}^\flat$ & $\mathrm{A}$ & $\mathrm{A}^\sharp/\mathrm{B}^\flat$ & $\mathrm{B}$\\
    \midrule
    $\ord(\mathrm{P})$ & 6 & 7 & 8 & 9 & 10 & 11\\
    \bottomrule
    \end{tabular}
    \caption{Pitches in an octave}
    \label{tab:octave}
\end{table}

A pitch in the pitch class $\mathrm P$ has a \textbf{pitch name} $\mathrm P_n$, where the subscript $n\in\Z$, called \textbf{octave number}, distinguishes frequencies in the same class $\mathcal P$. The difference in octave numbers relates to the frequency ratio in octaves between the pitches as
\begin{equation}
    \mathrm P_n:=\mathrm P_m\times 2^{n-m},
\end{equation}
then a pitch class $\mathrm C\in\mathcal P$ is composed of 
\begin{equation*}
    \mathrm{C}:=\{\cdots,\mathrm C_3,\mathrm C_4,\mathrm C_5,\cdots\}.
\end{equation*}
Within the same octave number $\mathcal P_n=\{\mathrm{C}_n,\cdots,\mathrm{B}_n\}$, pitches are connected by the distance in their ordinal numbers as
\begin{equation}
    \mathrm{P}_n:=\mathrm{Q}_n\times 2^{(\ord(\mathrm P) - \ord(\mathrm Q))/12}.
\end{equation}
According to this notation, any frequency corresponding to a pitch name $\mathrm{C}_4=\mathrm A_4\times2^{-3/4}\in\mathcal P_4$ can be determined by the reference frequency of \textit{one} pitch. $\mathrm A_4:=440\,\mathrm{Hz}$ is the most popular standard in the music community.

$\mathcal P$ is isomorphic to $\{0,1\}^{12}$ by a vectorization function $\bvec$ defined on any of its elements as 
\begin{equation*}
    \bvec(\mathrm P)_i=\bvec(\mathrm P_n)_i=\delta_{i,\ord(\mathrm P)}.
\end{equation*}
The collection of these vectors is a basis of $\R^{12}$.

The piano is an instrument that usually follows an equal temperament. Its keyboard is grouped by the octave number. Each group contains the keys corresponding to the pitches $\mathrm{C}_n$ to $\mathrm{B}_n$ with the same octave number, illustrated in \autoref{fig:transposition}.

\subsection{Symbolic music}

A piece of music is a time series. The simplest practice of composing music is picking pitches and deciding the key time points when they sound and are silent. These key time points usually follow some pattern called the \textbf{rhythm}. The rhythm allows for the recognition of the \textbf{beat} as a basic unit of musical time. A series of beats sets an integer grid on the time axis. A sound usually starts at some simple rational point in this grid, and its timespan, called the \textbf{value}, is also some simple fraction of one beat.

Thus, the temperament sets a coordinate system in the frequency domain, and the beat series sets a coordinate system in the time domain. On these coordinate systems, a segment with a pitch $\mathrm P_n$ on the frequency axis, a starting beat $b\in\Q$ and a value $v\in\Q$ along the time axis, is called a \textbf{note} $(\mathrm P_n, b, v)$, the information unit of music. A symbolic system, such as a music score, can record a piece of music with the symbols of the coordinate systems and the notes on it, which is the foundation of symbolic music. Once a reference frequency of one pitch is chosen, like $\mathrm A_4:=440\,\mathrm{Hz}$, the frequency of each pitch is determined. Once the length of the beats in the wall time, or the \textbf{tempo}, is chosen, the key time point of each sound is determined. More information for the music performance, such as timbres (spectrum signatures that feature an instrument), articulations (playing techniques), and dynamics (local treatments on the speed and volume), can also be annotated in the symbolic music. With these parameters, players with instruments or synthesizers lift the symbolic music to the playing audio.

\subsection{Chord}
The combination of multiple pitches forms a \textbf{chord}. By virtue of the octave equivalence, a chord $C$ can be simplified as a combination of pitch classes, namely $C\in 2^{\mathcal P}$. Then with the ordinal number $\ord(\cdot)$, every chord is naturally expressed as a set of number $\ord(C):=\{\ord(\mathrm{P}):\mathrm{P}\in C\}\subset\{0,\cdots,11\}$, or a binary vector $\bvec(C)=\bm c\in\{0,1\}^{12}$ where the entry is the value of an indicator function $c_i:=I_{\ord(C)}(i)$. Another equivalent expression is
\begin{equation*}
    \bvec(C)=\sum_{\mathrm P\in C}\bvec(\mathrm P).
\end{equation*}

The theory of harmony studies the interaction between pitches under a temperament. It reveals that different combinations have different musical timbres and harmonic brightness~\cite{laitz2012complete}. For example, a \textbf{C major} chord, $C_\text{C}=\{\mathrm C,\mathrm E,\mathrm G\}$, is bright and stable, while a \textbf{C minor} chord, $C_\text{Cm}=\{\mathrm C,\mathrm E^\flat,\mathrm G\}$, is dim and tense.

\subsubsection{Transposition}
\begin{figure}[!htbp]
    \centering
    \includegraphics[width=\linewidth]{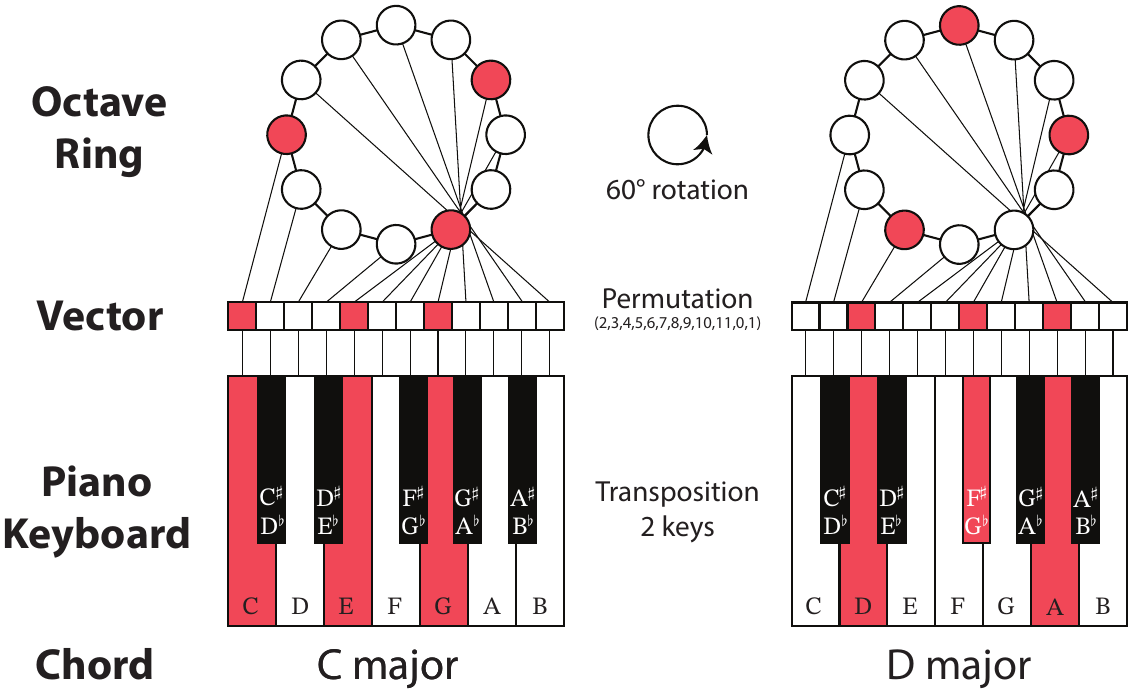}
    \caption{The example of transposition}
    \label{fig:transposition}
\end{figure}

We define the \textbf{transposition} operator $\mathcal T_i\, (i\in\Z)$. When it acts on a pitch,
\begin{equation*}
    \mathcal T_i(\mathrm P_n):=\mathrm P_n\times 2^{i/12}.
\end{equation*}
When it acts on a pitch class,
\begin{equation*}
    \mathcal T_i(\mathrm P):=\{\mathcal T_i(\mathrm P_n):n\in\Z\}.
\end{equation*}
In \autoref{fig:transposition}, we apply $\mathcal T_2$, a transposition of two semitones, to the C major chord $\{\mathrm C,\mathrm E,\mathrm G\}$, arriving at the chord $\{\mathrm D,\mathrm F^\sharp,\mathrm A\}$, which is called the D major chord in harmony theory. Harmony theory points out that a transposition doesn't change the \textbf{color} of a chord, but shifts the \textbf{tone} of a chord. Thus, the transposed chord always sounds similar to the original one but adapts to another set of pitches.

It's easy to verify
\begin{equation*}
    \ord(\mathcal T_i(\mathrm P))\equiv\ord(\mathrm P)+i\mod 12
\end{equation*}
Thus, for the action on $\mathcal P$
\begin{equation*}
    \mathcal T_i=\mathcal T_{i+12k},\quad k\in\Z
\end{equation*}
This means that $\mathcal T_1(\mathrm B)=\mathrm C$, linking the tail of an octave with the head. As shown at the top of \autoref{fig:transposition}, each pitch class can be mapped into an \textbf{octave ring}, where $\mathcal P$ forms a dodecagon inscribed in the circle, and $\mathcal T_i$ corresponds to the rotation by $i\times 30^\circ$, which leaves the dodecagon invariant. By checking the definition, $\{\mathcal T_0,\cdots,\mathcal T_{11}\}$ forms a group,
\begin{gather*}
    \mathcal T_0=\Id,\quad\mathcal T_i\mathcal T_j=\mathcal T_{i+j},\\
    \quad(\mathcal T_i\mathcal T_j)\mathcal T_k=\mathcal T_i(\mathcal T_j\mathcal T_k),\quad(\mathcal T_i)^{-1}=\mathcal T_{-i}.
\end{gather*}
This geometric intuition indicates that it is isomorphic to the cyclic group $\Z_{12}$ or the rotation group $C_{12}$. A transposition operator on $\mathcal P$ is equivalent to a group action of $C_{12}$ on $\mathcal P$.

When it acts on a chord, we define
\begin{equation*}
    \mathcal T_i(C)=\{\mathcal T_i(\mathrm P):\mathrm P\in C\}.
\end{equation*}
Thanks to the isomorphism $\bvec:\mathcal P\to\R^{12}$, there exists a group homomorphism $\mathbf D^\text{perm}|_{C_{12}}:C_{12}\to\mathsf{GL}(\R^{12})$ that satisfies
\begin{equation*}
    \bvec(\mathcal T_i(C)) = \mathbf D^\text{perm}(\mathcal T_i)\bvec(C),
\end{equation*}
and it's obvious that every $\mathbf D^\text{perm}(\mathcal T_i)$ is a permutation matrix. $\mathbf D^\text{perm}|_{C_{12}}$ is a permutation representation of $C_{12}$.

\subsubsection{Reflection}
We define the \textbf{reflection} operator $\mathcal R$. In a pitch class, it acts as
\begin{equation*}
    \ord(\mathcal R(\mathrm P))\equiv -\ord(\mathrm P)\mod 12,
\end{equation*}
And on a chord, it acts as
\begin{equation*}
    \mathcal R(C)=\{\mathcal R(\mathrm P):P\in C\}.
\end{equation*}
The behavior of $\mathcal R$ on the octave circle is a reflection w.r.t. the mirror passing through the $\mathrm C$ vertex and the $\mathrm F^\sharp/\mathrm G^\flat$ vertex of the inscribed dodecagon.
It is more general to think of the semidirect product $\{\mathcal T_0,\cdots,\mathcal T_{11}\}\rtimes\{\Id,\mathcal R\}$. Each transformation in the product with the form $\mathcal T_i\mathcal R$ is a reflection with a mirror either passing through two opposite vertices or passing through the middle points of two opposite edges of the inscribed dodecagon, leaving the dodecagon invariant. This geometry adopts dihedral group $D_{12}$ or the isomorphic $C_{12v}$. The transposition-reflection on $\mathcal P$ is equivalent to a group action of $D_{12}$ on $\mathcal P$.

Because $C_{12}$ is a subgroup of $D_{12}$, the previous representation $\mathbf D^\text{perm}|_{C_{12}}$ of $C_{12}$ induces $\mathbf D^\text{perm}:D_{12}\to\mathsf{GL}(\R^{12})$, which satisfies
\begin{equation*}
    \bvec(g.C) = \mathbf D^\text{perm}(g)\bvec(C)\quad \forall g\in D_{12}
\end{equation*}

\begin{figure}[!htbp]
    \centering
    \includegraphics[width=\linewidth]{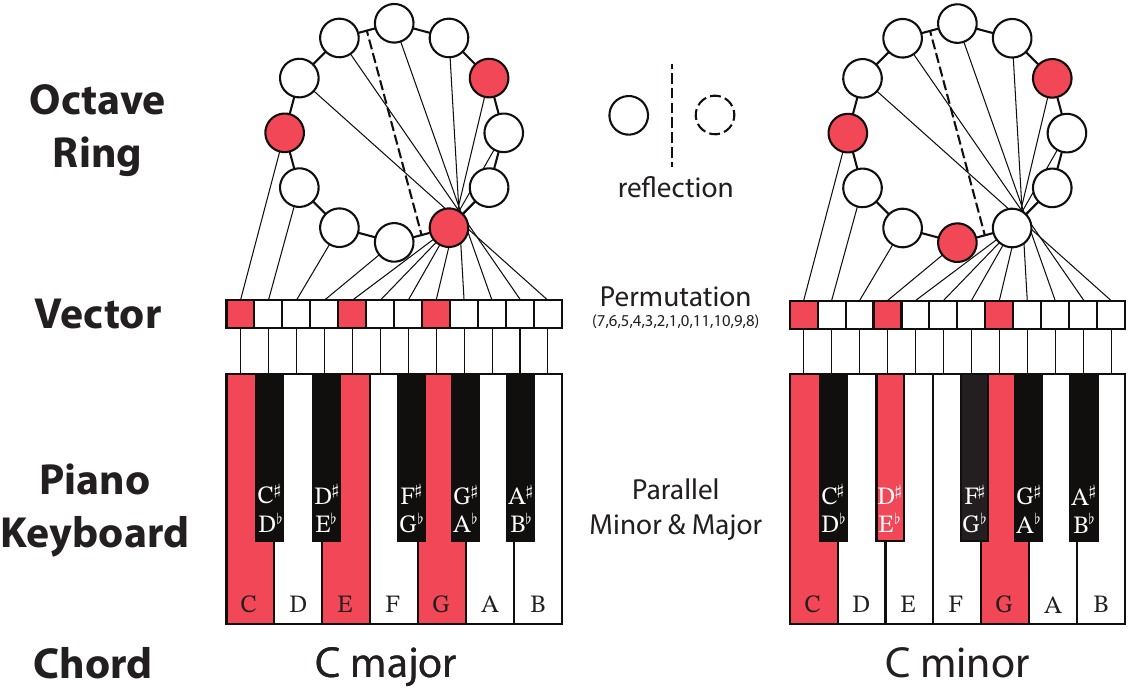}
    \caption{The example of reflection}
    \label{fig:reflection}
\end{figure}

In \autoref{fig:reflection}, we get $(\mathcal T_7\mathcal R)(C_\text{C})=C_\text{Cm}$. The underlying mirror is the dashed line. Harmony theory says that a reflection changes the \textbf{color} of a chord, switching between the bright, stable major chords and dim, tense minor chords, but if composed with an appropriate transposition, it keeps the \textbf{tone}.

\subsection{Homophony}
\textbf{Homophony} is a music composition framework, in which a music has a primary part called the \textbf{melody} and other auxiliary parts called the \textbf{accompaniment}. Most pop music follows it with a vocal melody and instrumental accompaniment. One of the most important accompaniment formats is the chord progression, a time series of chords with their starting beats and values $(C,b,v)$ \textit{with no timespan overlap}. It can be further detailed as a note series for performance by some \textbf{texture}, but the chord itself has included pivotal information to fulfill the musical requirement of the whole piece. The relationship of the melody notes $\{(\mathrm P_n,b,v)\}$ and the chord progression $\{(C,b,v)\}$ is restricted by the harmony theory as a probability distribution $P(\{(\mathrm P_n,b,v)\}|\{(C,b,v)\})$ like a generative model. For a simpler predictive model, it learns the map $\mathcal A$ from the melody to the best chord progression $\{(\mathrm P_n,b,v)\}\mapsto\{(C,b,v)\}$.

The transformation defined above should apply to the whole piece of music. Transposition on the melody $\{(\mathcal T_i\mathrm P_n,b,v)\}$ is the overall key shift, and its combination with reflection is involved in the major-minor variation. As a result, the harmony restriction requires that the accompaniment, especially the chord progression, should transform accordingly. For a generative model, this implies an equivariant distribution
{\footnotesize
\begin{equation*}
    P(\{(g.\mathrm P_n,b,v)\}|\{(g.C,b,v)\})=P(\{(\mathrm P_n,b,v)\}|\{(C,b,v)\}).
\end{equation*}}
For a predictive model, it needs to learn an equivariant function where $\mathcal Ag=g.\mathcal A$ holds.

\section{Method}

\subsection{Embedding music into vectors}
A minimum value of $u$ is used as a time step. The melody notes $\mathcal N=\{(\mathrm P_n,b,v)\}$ are embedded as a series of vectors $\bm m^{(k)}\in [0,1]^{12}$, where $\bm m^{(k)}$ records the sounding notes during the timespan between $(k-1)u$ and $ku$. Inspired by \cite{lin2017automatic}, the embedding builds on the relative contribution of each note during the timespan.
\begin{equation*}
\footnotesize
    \bm m^{(k)} = \sum_{(\mathrm P_n,b,v)\in\mathcal N}\frac{\max\{ku,b+v\} - \min\{(k-1)u, b\}}{u}\cdot\bvec(P_n)
\end{equation*}

The chord progression $\mathcal C=\{(C,b,v)\}$ can be similarly embedded as vectors $\bm c^{(k)}$; however, the chord progression is much sparser than the melody notes. If $u$ is small enough (e.g. half beat), $u$ will be the common factor of every $b$ and $b+v$, thus the term $\frac{\max\{ku,b+v\} - \min\{(k-1)u, b\}}{u}$ will be binary. And because of no timespan overlap in the chord progression, each summation has exactly one term. In this case, $\bm c^{(k)}\in\{0,1\}^{12}$ itself is the vectorized form of \textit{one} chord, that is, the chord state during the time between $(k-1)u$ and $ku$. The collection of $\bm c^{(k)}$ describes a state transition series.

This featurization naturally pulls the $D_{12}$ group action on pitch classes back to the permutation representation $\mathrm D^\text{perm}$ on $\R^{12}$. If there are totally $T$ time steps, we collect the melody vectors of $\mathcal N=\{(\mathrm P_n,b,v)\}$ as a matrix $\mathbf M=(\bm m^{(1)},\cdots,\bm m^{(T)})\in\R^{12\times T}$ and the chord vectors of $\mathcal C=\{(C,b,v)\}$ as a matrix $\mathbf C=(\bm c^{(1)},\cdots,\bm c^{(T)})\in\R^{12\times T}$, then the featurization of $\mathcal N=\{(g.\mathrm P_n,b,v)\},\mathcal C=\{(g.C,b,v)\}$ is $\mathbf D^\text{perm}(g)\mathbf M,\mathbf D^\text{perm}(g)\mathbf C$ for all $g\in D_{12}$.

We aim to build a predictive model $\mathcal A$ which takes in $\mathbf M$ and predicts $\mathbf C$. Hence, its equivariant condition is 
\begin{equation}\label{eq:cond}
    \mathcal A\mathbf D^\text{perm}=\mathbf D^\text{perm}\mathcal A.
\end{equation}

\subsection{Decomposition of the permutation representation}

The first step of constructing and parametrizing an expressive neural network $\mathcal A$ that satisfies the equivariant condition \autoref{eq:cond} is building the equivariant linear map. One common practice is transforming the space of the permutation representation into irreducible representations of $D_{12}$, and designing each linear layer in each irreducible representation channel~\cite{geiger2022e3nn}. For a $l_a$-dimension irreducible representation indexed by $a$, if
\begin{equation*}
    \mathbf H^{(a)}\in\R^{l_a\times s}=\begin{pmatrix}\bm h_1^{(a)} & \cdots & \bm h_s^{(a)}\end{pmatrix}
\end{equation*}
composed of column vectors in $a$-channel is a matrix in $a$-channel \textit{with multiplicity} $s$. Then for each $a$-channel vector $\bm h^{(a)}$, the weight of the equivariant linear layer $\bm W^{(ab)}\in\R^{l_a\times l_b}$ working between the channel $a$ should keep the equivariance $\mathbf W^{(ab)}\mathbf D^{(a)}\bm h^{(a)} = \mathbf D^{(b)}\mathbf W^{(ab)}\bm h^{(a)}$ between two channels. According to Schur's lemma~\cite{serre1977linear}, $\mathbf W=\mathbf I_{l_a\times l_a}\delta_{ab}$ is the only solution to this condition. Therefore, the general linear layer is only within the same channel working on the mixing multiplicities, parametrized as a learnable weight $\mathbf W^{(a)}\in\R^{s_\text{in}\times s_\text{out}}$ that maps $\mathbf H^{(a)}\in\R^{l_a\times s_\text{in}}$ to $\mathbf H^{(a)}\mathbf W\in\R^{l_a\times s_\text{out}}$.

\begin{figure}[!htbp]
    \centering
    \includegraphics[width=\linewidth]{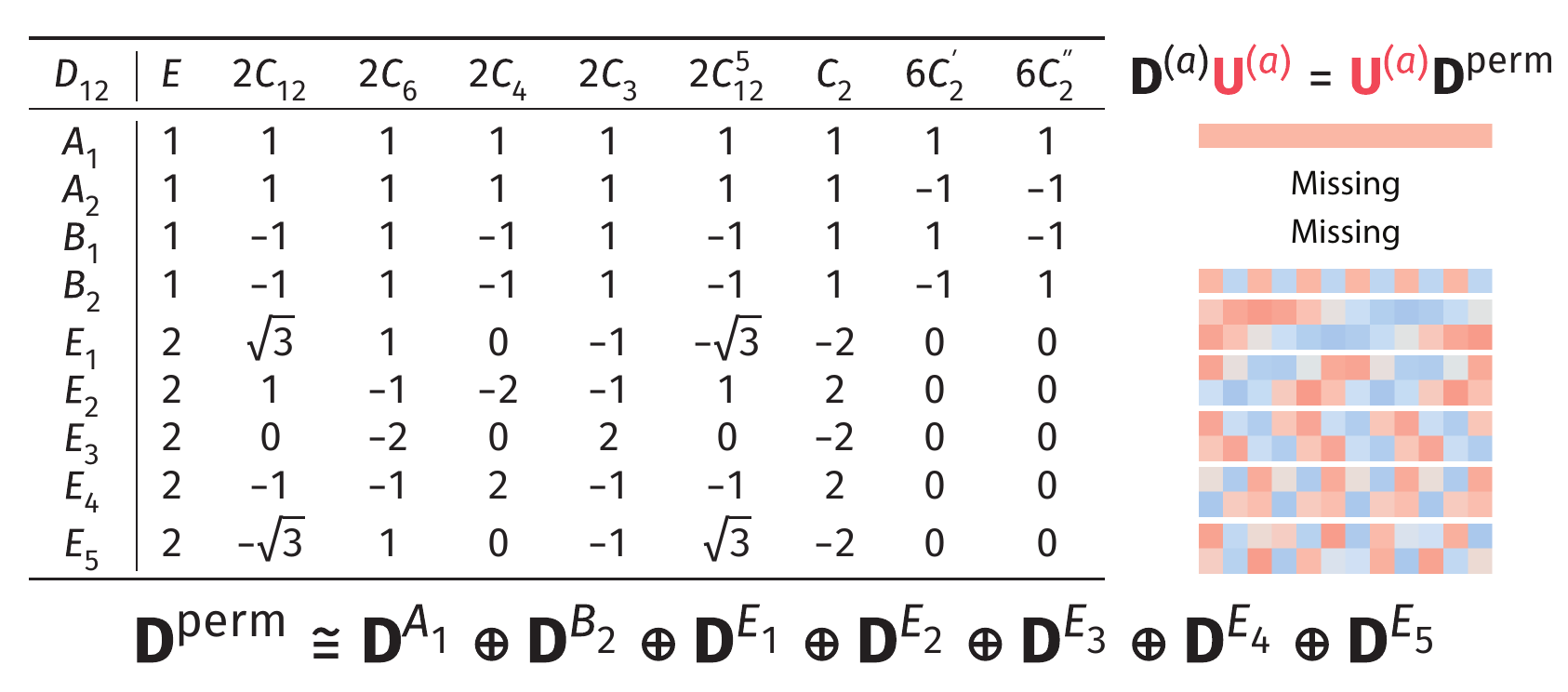}
    \caption{The character table of $D_{12}$, the canonical decomposition of its permutation representation, and the illustration of change of basis matrices.}
    \label{fig:char}
\end{figure}

Specifically, $\mathbf D^\text{perm}$ is a 12-dimensional reducible representation of $D_{12}$. According to the character table (\autoref{fig:char}), it can be decomposed into the direct sum of 7 distinct irreducible representations, where the other two components transformed as $A_2$ or $B_1$ vanish. Because the non-zero coefficients in the canonical decomposition are all one, the solution of the change of basis matrix $\mathbf U^{(a)}\in\R^{l^{(a)}\times 12}$ via 
\begin{equation}\label{eq:cob}
    \mathbf D^{(a)}(g)\mathbf U^{(a)}=\mathbf U^{(a)}\mathbf D^\text{perm}(g)
\end{equation}
is unique up to an isomorphism. For the sake of a simple and stable neural network training, we have $\mathbf U^{(a)}$ to be
\begin{itemize}
    \item Real: This possibility is guaranteed because all the irreducible representations of $D_{12}$ are real inherently.
    \item Orthogonal: $\mathbf U^{(a)} (\mathbf U^{(a)})^\top=\mathbf I_{l^{(a)}\times l^{(a)}}$. This can be realized by normalizing the solution.
\end{itemize}
Then each $\mathbf U^{(a)}$ has its pre-determined value stored as a constant before invoking the neural network. As illustrated in \autoref{fig:char}, different change of basis matrices capture different features in the octave. For example, $\mathbf D^{A_1}$ extracts the average signal over the octave, $\mathbf D^{B_2}$ compares the difference in minor seconds, while $\mathbf D^{E_3}$ mixes the patterns in major thirds.

\subsection{An overview of the Music10x model}
\begin{figure}[!htbp]
    \centering
    \includegraphics[width=\linewidth]{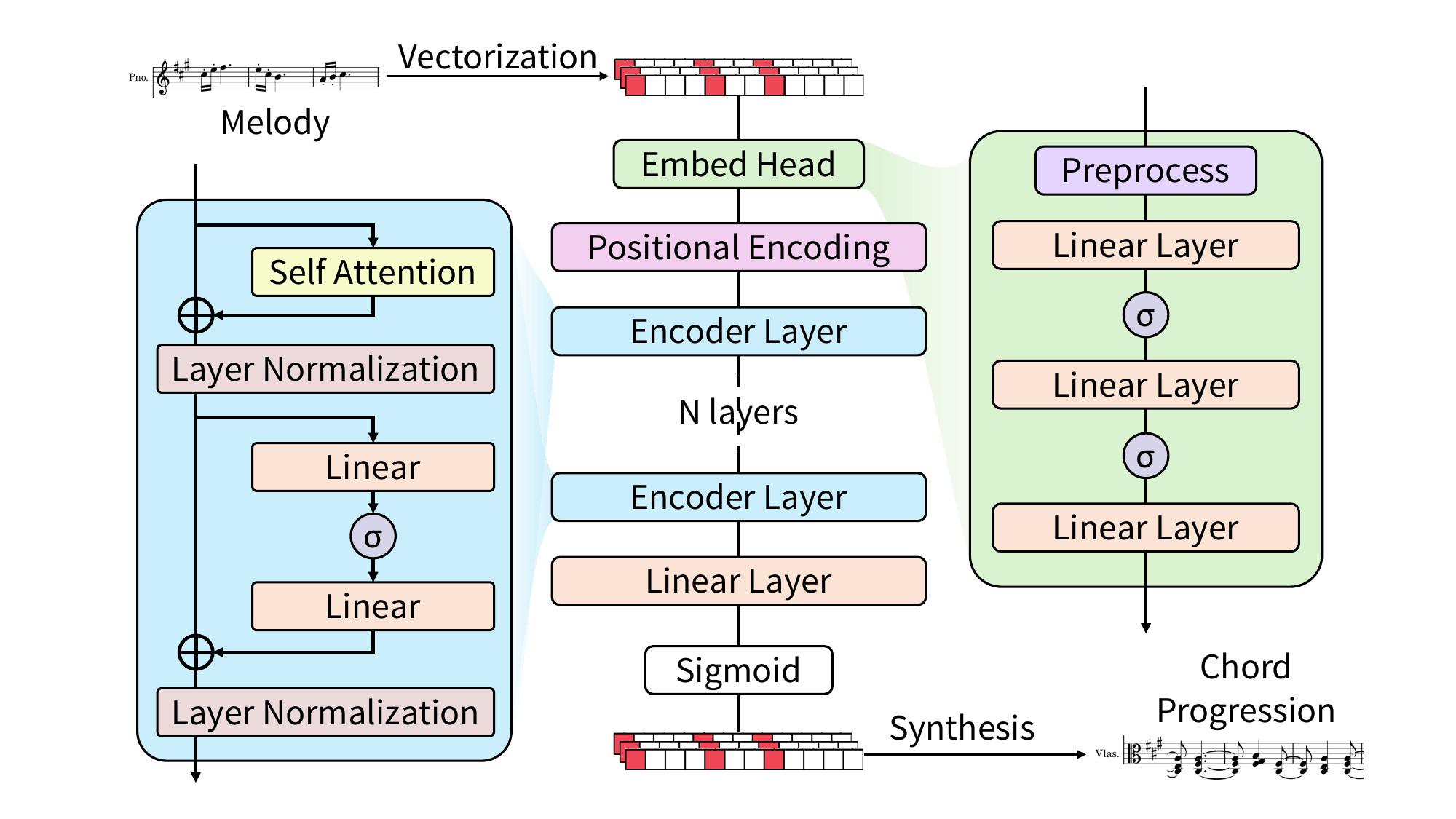}
    \caption{Music10x backbone}
    \label{fig:arch}
\end{figure}

As illustrated in \autoref{fig:arch}, Music101 and Music102 share the same backbone. The preprocessing layer in Music101 is an identity map, and the other layers follow the traditional implementation in a transformer \cite{Transformer}. As a result, Music101 doesn't naturally satisfy the equivariant condition \autoref{eq:cond}.

In Music102, the linear layer, the positional encoding, the self-attention layer, the layer normalization, and the non-linearity $\sigma$ are reformulated to be equivariant, as detailed in the following section.

\subsection{Other $D_{12}$-equivariant layers}

\textbf{$D_{12}$-preprocessing}\quad
For each column vector $\bm m$ in $\mathbf M$ which transforms as $\mathbf D^\text{perm}$, a $D_{12}$-featurization layer pushes it forward to a representation vector $\bm h^{(a)}\in\R^{l^{(a)}}$ \textit{belongs to} $a$-\textit{channel} by
\begin{equation*}
    \bm h^{(a)}=\mathbf U^{(a)}(\bm m+b_a\bm 1_{12})
\end{equation*}
where $b_a$ is a learnable scalar and $\bm 1_{12}\in\R^{12}$ is a vector of ones. We can check that
\begin{align*}
    &\mathbf U^{(a)}\cdot(\mathbf D^\text{perm}(g)\bm m+b_a\bm 1_{12})\\
    =&\mathbf U^{(a)}\mathbf D^\text{perm}(g)\cdot(\bm m+b_a\bm 1_{12})\\
    =&\mathbf D^{(a)}(g)\mathbf U^{(a)}\cdot(\bm m+b_a\bm 1_{12})
\end{align*}
thus $\bm h^{(a)}$ from the $D_{12}$-featurization layer transforms as $\mathbf D^{(a)}$.

\hspace{1pt}

\noindent\textbf{$D_{12}$-equivariant activation function}\quad
Common activation functions, like ReLU or Sigmoid, don't behave equivariantly between linear layers. However, if a vector $\bm h=(h_1,\cdots,h_{12})^\top$ transforms as a $\mathbf D^\text{perm}$ that maps the $i$-th element to $p(i)$-th one, for any element-wise function $f(\cdot)$, we have
\begin{equation}\label{eq:perm}
    \begin{aligned}
    f(\mathbf D^\text{perm}\bm h) &= \begin{pmatrix}
        f(h_{p(1)})\\
        \cdots\\
        f(h_{p(12)})
    \end{pmatrix}=\mathbf D^\text{perm}\begin{pmatrix}
        f(h_{1})\\
        \cdots\\
        f(h_{12})
    \end{pmatrix}\\&=\mathbf D^\text{perm}f(\bm h)
    \end{aligned}
\end{equation}
that is, \textit{any element-wise function commutes with the permutation representation}, including any common non-linearity we may apply.

Take the transposition of the \autoref{eq:cob}, it holds for any $g\in D_{12}$ that
\begin{equation*}
    (\mathbf U^{(a)})^\top(\mathbf D^\text{perm}(g))^\top=(\mathbf U^{(a)})^\top(\mathbf D^{(a)}(g))^\top
\end{equation*}
It is possible to have both $\mathbf D^\text{perm}$ and $\mathbf D^{(a)}(g)$ orthogonal representations, which leads to
\begin{equation*}
    (\mathbf U^{(a)})^\top(\mathbf D^\text{perm}(g^{-1}))=(\mathbf U^{(a)})^\top(\mathbf D^{(a)}(g^{-1}))
\end{equation*}
for any $g\in D_{12}$. Thus $(\mathbf U^{(a)})^\top$ is the matrix that pulls the $a$-channel vector back to the permutation representation. Then a non-linearity $\sigma(\cdot)$ can be incorporated by $\bm h^{(a)}\mapsto\mathbf U^{(a)}\sigma((\mathbf U^{(a)})^\top\bm h^{(a)})$ in $a$-channel, because
\begin{align*}
    &\mathbf U^{(a)}\sigma((\mathbf U^{(a)})^\top\mathbf D^{(a)}(g)\bm h^{(a)})\\=&\mathbf U^{(a)}\sigma(\mathbf D^\text{perm}(g)\cdot(\mathbf U^{(a)})^\top\bm h^{(a)})\\
=& \mathbf U^{(a)}\mathbf D^\text{perm}(g)\sigma((\mathbf U^{(a)})^\top\bm h^{(a)})\\
=&\mathbf D^{(a)}(g)\mathbf U^{(a)}\sigma((\mathbf U^{(a)})^\top\bm h^{(a)})
\end{align*}

\hspace{1pt}

\noindent\textbf{$D_{12}$-equivariant positional encoding}\quad
The positional encoding is added to the sequential features to make the self-attention mechanism position-aware. For a sequence of $d$-dimensional word embedding $\mathbf X\in\R^{L\times d}$, the encoded sequence is
\begin{gather*}
    \mathop\mathrm{PE}(\mathbf X)_{t,k}=X_{t,k} +S_{t,k},\\  S_{t,k}:= \begin{cases}
    \sin(w_it) & k=2i\\
    \cos(w_it)     & k=2i+1
    \end{cases},\\w_i:=\frac{1}{10000^\frac{2i}{d}},\  i=0,1,\cdots,d/2-1.
\end{gather*}
In our $D_{12}$-equivariant architecture, the sequence in $a$-channel with multiplicity $s_a$ is a tensor $\mathbf X^{(a)}\in\R^{L\times l_a\times s_a}$, where $\mathbf X^{(a)}_{t,\cdot,\cdot}\in\R^{l_a\times s_a}$ is a matrix in $a$-channel. We define the positional encoding for $a$-channel as
\begin{gather*}
    \mathop\mathrm{PE}(\mathbf X^{(a)})_{t,\cdot,k} = \mathbf X^{(a)}_{t,\cdot,k} + (\mathbf U^{(a)}\mathbf S_{t,\cdot,\cdot})_{\cdot,k},\\ \mathbf S\in\R^{L\times 12\times d},\\ \mathbf S_{t,\cdot,k}\equiv S_{t,k}
\end{gather*}
Because $\mathbf S$ is a constant along the representation dimension, the permutation acts trivially $\mathbf D^\text{perm}(g)\mathbf S_{t,\cdot,\cdot}=\mathbf S_{t,\cdot,\cdot}$. It follows that
\begin{align*}
   & \mathop\mathrm{PE}(\mathbf D^{(a)}(g)\mathbf X^{(a)}_{t,\cdot,\cdot})
\\=&\mathbf D^{(a)}(g)\mathbf X^{(a)}_{t,\cdot,\cdot}+\mathbf U^{(a)}\mathbf S_{t,\cdot,\cdot}\\
=&\mathbf D^{(a)}(g)\mathbf X^{(a)}_{t,\cdot,\cdot}+\mathbf U^{(a)}\mathbf D^\text{perm}(g)\mathbf S_{t,\cdot,\cdot}\\
=&\mathbf D^{(a)}(g)\mathbf X^{(a)}_{t,\cdot,\cdot}+\mathbf D^{(a)}(g)\mathbf U^{(a)}\mathbf S_{t,\cdot,\cdot}\\
=& \mathbf D^{(a)}(g)(\mathbf X^{(a)}_{t,\cdot,\cdot}+\mathbf U^{(a)}\mathbf S_{t,\cdot,\cdot})\\
=& \mathbf D^{(a)}(g)\mathop\mathrm{PE}(\mathbf D^{(a)}(g)\mathbf X^{(a)}_{t,\cdot,\cdot}).
\end{align*}
Thus, the positional encoding is equivariant.

\hspace{1pt}

\noindent\textbf{$D_{12}$-equivariant multihead self-attention}\quad
Through $D_{12}$-equivariant linear layers, the sequence $\mathbf X^{(a)}$ is transformed into the query sequence $\mathbf Q^{(a)}$, the key sequence $\mathbf K^{(a)}$, and the value sequence $\mathbf V^{(a)}$ in each channel. We split the multiplicity of sequences $\mathbf Q^{(a)}$, $\mathbf K^{(a)}$ into $N_h$ heads and concatenate different channels as
\begin{equation*}
    \mathbf Y_{t,\cdot}=\bigoplus_a \vvec(\mathbf Y^{(a)}_{t,\cdot,\cdot}),\quad \vvec(\mathbf Y^{(a)}_{t,\cdot,\cdot}) = \begin{pmatrix}
        \mathbf Y^{(a)}_{t,\cdot,1}\\
        \vdots\\
        \mathbf Y^{(a)}_{t,\cdot,s_a/N_h}
    \end{pmatrix}
\end{equation*}. Obviously the concatenated sequence transforms as the representation $\mathbf D=\bigoplus_a\bigoplus_{i=1}^{s_a/N_h}\mathbf D^{(a)}$, which is orthogonal when all the $\mathbf D^{(a)}$ is orthogonal. When this concatenation applies to $\mathbf Q^{(a)},\mathbf K^{(a)}$ and results in $\mathbf Q,\mathbf K\in\R^{L\times\sum_{a}s_a}$, the attention weight $\mathbf\alpha=\sigma(\frac{\mathbf Q\mathbf K^\top}{\sqrt d})\in\R^{L\times L}$ becomes invariant by virtue of the orthogonality of $\mathbf D$ because
\begin{equation*}
    \mathbf Q\mathbf D(g)(\mathbf K\mathbf D(g))^\top=\mathbf Q\mathbf D(g)(\mathbf D(g))^\top\mathbf K^\top = \mathbf Q\mathbf K^\top
\end{equation*}
The weighted sum $\mathbf\alpha\mathbf V^{(a)}$ naturally transforms as a representation which $\mathbf V^{(a)}$ follows.

This mechanism also applies to multi-head attention if each head's queries and keys follow the same orthogonal representation.

\hspace{1pt}

\noindent\textbf{$D_{12}$-equivariant layer normalization}\quad
Taking advantage of the \autoref{eq:perm} and that the permutation doesn't affect the vector elements' variance and mean, we define the layer normalization as
\begin{gather*}
    \mathop\mathrm{LN}(\mathbf X^{(a)})_{t,\cdot,\cdot} = \mathbf U^{(a)}\mathbf Z,\\
    \mathbf Z_{t,\cdot,k}=\left(\frac{\tilde{\mathbf X}^{(a)}_{t,\cdot,\cdot}-\mu(\tilde{\mathbf X}^{(a)}_{t,\cdot,\cdot})}{\sqrt{\mathop\mathrm{Var}(\tilde{\mathbf X}^{(a)}_{t,\cdot,\cdot})+\epsilon}}\right)_{\cdot,k}\cdot\gamma_k^{(a)}+\beta_k^{(a)},\\
    \tilde{\mathbf X}^{(a)}_{t,\cdot,\cdot}=(\mathbf U^{(a)})^\top\mathbf X^{(a)}_{t,\cdot,\cdot}.
\end{gather*}
where $\mathop\mathrm{Var}(\cdot)$ takes the variance of all elements of the input, $\mu(\cdot)$ takes the mean, $\epsilon$ is a small positive number for numerical stability, and $\bm\gamma^{(a)},\bm\beta^{(a)}\in\R^{s_a}$ are learnable parameters.

\subsection{Loss function}
A weighted binary cross-entropy loss is defined between the predicted $\tilde{\mathbf C}$, whose each column is the logit of the 12 pitches, and the ground truth $\mathbf C$.
\begin{equation*}
    \mathcal L(\tilde{\mathbf C};\mathbf C)=\sum_{i,j}-w_j(C_{ij}\log\tilde C_{ij}+(1-C_{ij})\log(1-\tilde C_{ij})).
\end{equation*}
The weights emphasize the chord transition point by
\begin{equation*}
    w_j=\begin{cases}
        1 & \mathbf C_{\cdot,j}=\mathbf C_{\cdot,j-1}\\
        2 & \mathbf C_{\cdot,j}\ne\mathbf C_{\cdot,j-1}\ \text{or}\ j=1\\
    \end{cases}.
\end{equation*}

\section{Experiments}
The code conducting the experiments can be found in \href{https://github.com/Benzoin96485/music102}{our Github repo}.

\subsection{Data Acquisition and processing}
The model is trained on the POP909 Dataset \cite{Wang2020POP909AP}, which contains 909 pieces of Chinese pop music with the melody stored in MIDI and the chord progression annotation stored as a time series. We extract the melody from the MIDI file of each song and match it with the chord progressions and beat annotations using the time stamps. The minimal $u$ is set to be 1/2 beat in our experiments. The dataset is randomly split into 707 pieces in the training set, 100 pieces in the validation set, and 100 pieces in the test set. The vectorization and the loss weights are precomputed before training.

\subsection{Comments on the numerical stability}
We tried another flavor of equivariant non-linearities and layer normalizations adopted in Equiformer \cite{liao2022equiformer} and SE(3)-transformer \cite{fuchs2020se}, the norm-gated ones. For instance, the non-linearity $\sigma(\cdot)$ acting on an equivariant vector $\bm v$ can remain equivariant by the invariance of the norm under unitary representations.
\begin{equation*}
    \tilde\sigma(\bm v)=\sigma(\|\bm v\|)\cdot\frac{\bm v}{\|\bm v\|}.
\end{equation*}
However, due to the sparsity of the input melody (a large number of zero column vectors in $\mathbf M$), the norm at the denominator induces unavoidable gradient explosion, which usually halts the training after several batches or epochs, even when combined with a small constant bias or trainable equivariant biases. This is one of the motivations during the experiment of shifting the non-linearity, the positional encoding, and the layer normalization to a vector transforming as the permutation representation by \autoref{eq:perm}. It is the key observation that enables the whole experiment after dozens of modified versions of the norm-gated flavor.

\subsection{Music synthesis and auditory examples}
The output $\tilde{\mathbf C}$ is rounded as a binary $\hat{\mathbf C}$ with a cutoff of logit 0.5. Mapping $\hat{\mathbf C}$ back to the pitch classes, MuseScore4 synthesizes the input melody and the output chord progression simultaneously to be a piece of complete music. Audio examples from Music102 output on the test set, as well as transposed versions of one piece to demonstrate its equivariance, can be found in \href{https://github.com/Benzoin96485/music102}{our Github repo}.

\subsection{Comparison between Music101 and Music102}

Limited by the computation resource and time, the hyperparameters haven't been thoroughly scanned but locally searched on several key components, including the length of features, the number of encoding layers, and the learning rate, arriving at a suboptimal amount of parameters.

In addition to the loss itself, the other two metrics also evaluate the performance of the model in reproducing the chord progression label. The cosine similarity is defined as
\begin{equation*}
    \mathop{\mathrm{CosSim}}(\tilde{\mathbf C};\mathbf C)=\frac{1}{L}\sum_j\frac{\hat {\mathbf C}_{\cdot,j}^\top\mathbf C_{\cdot,j}}{\|\hat{\mathbf C}_{\cdot,j}\|\|\mathbf C_{\cdot,j}\|}.
\end{equation*}
The exact accuracy is defined as
\begin{equation*}
    \mathop{\mathrm{Acc}}(\tilde{\mathbf C};\mathbf C)=\frac{1}{L}\sum_jI_{\hat {\mathbf C}_{\cdot,j}, \mathbf C_{\cdot,j}}.
\end{equation*}
As shown in \autoref{tab:perf}\,($\downarrow$ means the lower the better, $\uparrow$ means the higher the better), Music102 reaches a better performance with a less amount of parameters. This indicates the potential data and computational efficiency of Music102 when the model performance scales with larger parameters and data.
\begin{table}[!htbp]
    \centering
    \tiny
    \begin{tabular}{c|cc}
    \toprule
    Model (Amount of Params$\downarrow$)     &  Music102 (760030) & Music101 (6850060)\\
    \midrule
    Weighted BCE loss ($\downarrow$)   & 0.5652 & 0.5807\\
    Cosine similarity ($\uparrow$)   & 0.6727 & 0.6638\\
    Exact accuracy ($\uparrow$)   & 0.1783 & 0.1141\\
    \bottomrule
    \end{tabular}
    \caption{Performance comparison among Music10x}
    \label{tab:perf}
\end{table}

\section{Conclusion}
To the best of our knowledge, this is the first transformer-based seq2seq model that considers the word-wise symmetry in the input and output word embeddings. As a result, the universal schemes in the transformer for natural language processing, including layer normalization and positional encoding, need to be adapted to this new domain. Our modification from a traditional transformer Music101 to the fully equivariant Music102 without essential backbone changes shows that there are out-of-the-box equivariant substitutions for a self-attention-based sequence model. Taking full advantage of the property of the permutation representation, we explore a more flexible and stable framework of equivariant neural networks on a discrete group.

Given the efficiency and accuracy of the chord progression task, we expect that Music102 could be a general backbone of the equivariant music generation model in the future. We believe that the mathematical structure within the music provides further implications for computational music composing and analysis.

\begin{acknowledgments}
The author thanks Yucheng Shang and Weize Yuan for their endeavor on the Music101 prototype.
\end{acknowledgments} 

\bibliography{icmc2025_paper_template}

\end{document}